\documentclass[prl,amsmath,twocolumn,showpacs,superscriptaddress]{revtex4-1}

\usepackage{graphicx}
\usepackage{subfig}	 
\usepackage{psfrag}
\usepackage{multirow}
\usepackage{natbib}
\usepackage{color}
\usepackage{wrapfig}
\usepackage[english]{babel} 
\usepackage{amsmath,amsthm,amssymb,amsfonts,latexsym} 
\usepackage[latin1]{inputenc} 
\newlength{\picwidth}
\pdfpageattr {/Group << /S /Transparency /I true /CS /DeviceRGB>>}  

\newcommand{\beq}{\begin{equation}}
\newcommand{\eeq}{\end{equation}}
\newcommand{\ben}{\begin{eqnarray}}
\newcommand{\een}{\end{eqnarray}}
\newcommand{\bea}{\begin{array}}
\newcommand{\eea}{\end{array}}

\newcommand{\bef}{\begin{figure}}
\newcommand{\eef}{\end{figure}}

\begin{document}

\title{Oxygen defects in phosphorene}

\author{A. Ziletti }

\affiliation{Department of Chemistry, Boston University, 590 Commonwealth Avenue, Boston Massachusetts 02215, USA}
\author{A. Carvalho}
\affiliation{Graphene Research Centre and Department of Physics, National University of Singapore, 117542, Singapore}

\author{D. K. Campbell}
\affiliation{Department of Physics, Boston University, 590 Commonwealth Avenue, Boston Massachusetts 02215, USA}

\author{D. F. Coker}
\affiliation{Department of Chemistry, Boston University, 590 Commonwealth Avenue, Boston Massachusetts 02215, USA}
\affiliation{Freiburg Institute for Advanced Studies (FRIAS), University of Freiburg, D-79104, Freiburg, Germany}

\author{A. H. Castro Neto}
\affiliation{Graphene Research Centre and Department of Physics, National University of Singapore, 117542, Singapore}
\affiliation{Department of Physics, Boston University, 590 Commonwealth Avenue, Boston Massachusetts 02215, USA}

\begin{abstract}
Surface reactions with oxygen are a fundamental cause of the degradation
of phosphorene. 
Using first-principles calculations, we show that for each oxygen atom adsorbed onto phosphorene there is an energy release
of about 2 eV. Although the most stable oxygen adsorbed forms are electrically inactive and lead only to minor
distortions of the lattice, there are low energy metastable forms which introduce deep donor and/or acceptor levels in the gap.
We also propose a mechanism for phosphorene oxidation and we suggest that dangling oxygen atoms increase the hydrophilicity of phosphorene. \\ 
\end{abstract}
\pacs{73.20.At,73.20.Hb}
\maketitle


Phosphorene, a single layer of black phosphorus\cite{rodin2014,liu2014}, has revealed extraordinary
functional properties which make it a promising material not only for
exploring novel physical phenomena but also for practical applications.
In contrast to graphene, which is a semi-metal, phosphorene is a semiconductor with a quasiparticle band gap of 2 eV. The optical band gap is reduced to 1.2~eV, because of the large exciton binding energy (800 meV)\cite{tran2014,rodin2014b}. 
Phosphorene's peculiar structure of parallel zig-zag rows leads to very anisotropic electron and hole masses, optical absorption and mobility\cite{low2014,xia2014,qiao2014}.
Both its gap and the effective masses can be tuned by
stressing phosphorene's naturally pliable waved structure.
Strain along the zigzag direction can switch the gap between direct and indirect\cite{fei2014}
and compression along the direction perpendicular to the layers
can in principle even transform the material into a metal or semimetal\cite{rodin2014}.
Phosphorene has also scored well as a functional material for
two-dimensional electronic and optoelectronic devices.
Multi-layer phosphorene field effect transistors have already
been demonstrated to exhibit on-off current ratios exceeding 10$^5$,
field-effect mobilities of 1000 cm$^2$/Vs\cite{li2014},
and fast and broadband photodetection\cite{buscema2014}.

An invariable issue encountered in the manipulation of phosphorene is the control of the oxidation.
The presence of exposed lone pairs at the surface makes phosphorus  very reactive to air. Surface oxidation is made apparent by the roughening, which grows exponentially during the first hour after exfoliation\cite{koenig2014}
and contributes to increasing contact resistance, lower carrier mobility
and possibly to the mechanical degradation and breakdown.
Thus, identifying the mechanisms of phospherene oxidation -- including the electrically active forms of oxygen
and how they are introduced -- is essential to understanding the real material and its applications.

In this Letter, we show that oxygen chemisorption onto phosphorene is exoenergetic
and leads to the formation of neutral defects,
as well as to metastable electrically active defect forms.
We also discuss the conditions necessary for extensive oxidation 
and propose strategies to control it. 

Oxygen defects were modeled using first-principles calculations
based on density functional theory (DFT), as implemented in the 
{\sc Quantum ESPRESSO} package\cite{qe}. 
We used three different approximations for the exchange-correlation energy: 
the semilocal generalized gradient approximations PBE\cite{pbe} and PBEsol\cite{pbesol} functionals, 
and the HSE06\cite{hse06} range-separated hybrid functional. 
The band structures were computed with the PBEsol functional because it has proven superior in determining lattice parameters of solids\cite{pbesol}. 
On the other hand, we validated the binding energies using the PBE  and HSE functionals, 
the former being typically superior to PBEsol for dissociation or cohesive energies\cite{pbesol},
and the latter to account for the role of exchange in determining the oxidation states of the oxygen molecule\cite{colleoni2014}.
Unless otherwise stated, all energies reported were obtained with the PBE functional. Activation energies were obtained using the climbing nudged elastic
band (NEB) method\cite{neb}. Vibrational frequencies of oxygen defects were calculated within the framework of density functional perturbation theory\cite{baroni2001}. Additional computational details are presented in \emph{Supplemental Material}.

Although the fast degradation of black phosphorus when exposed to air\cite{brunner1979,yau1992,koenig2014,gomez2014,farnsworth2014}
may involve the reaction with molecules other than oxygen (e.g. water vapor),
here we first consider, for simplicity, only the adsorption and incorporation of oxygen.
In particular, it is crucial to determine whether O chemisorption is energetically favored or not. 
To this end, for each oxidized structure we calculate the average binding energy for an oxygen atom, $E_{b}$, defined as
$E_{b}=-1/N_{\rm O} \left[ E_{ox}-\left(E_{p}+N_{\rm O}E_{\rm O_2}/2 \right) \right]
\label{eq:e-bind}$
where $N_{\rm O}$ is the number of O atoms in the cell used in the calculation, $E_{ox}$, $E_{p}$ and $E_{\rm O_2}$ are the total energies of the oxidized phosphorene, 
the pristine phosphorene, and  the ${\rm O}_2$ (triplet) molecule, respectively. 
According to the definition above, a positive $E_{b}$ indicates that the chemisorption is exothermic (energetically favored).

Chemisorbed or interstitial oxygen atoms can occupy numerous positions in the phosphorene lattice.
The most relevant structures are listed on Table \ref{table:e-bind}. 
These are also depicted in Figs.~1-3.
\begin{table}[ht]
\caption{\small Binding energies for the isolated oxygen impurities.} 
\centering 
\begin{tabular}{l   c  c c } 
\hline\hline 
 \multirow{2}{*}{Structure} &  $E_{b}$(eV) & \\
 & PBE & PBEsol \\
 \hline 
Dangling  &  2.08 &2.22 \\ 
Interstitial bridge  & 1.66 & 1.77 \\ 
Horizontal bridge  & $-$0.01 & 0.21 \\ 
Diagonal bridge    & $-$0.08 & 0.22  \\  
\hline\hline 
\end{tabular}
\label{table:e-bind}
\end{table}

\begin{figure}[htb]
\centering
  \subfloat[]{
    \includegraphics*[trim=0pt -45pt 0pt 0pt, width=2.8cm]{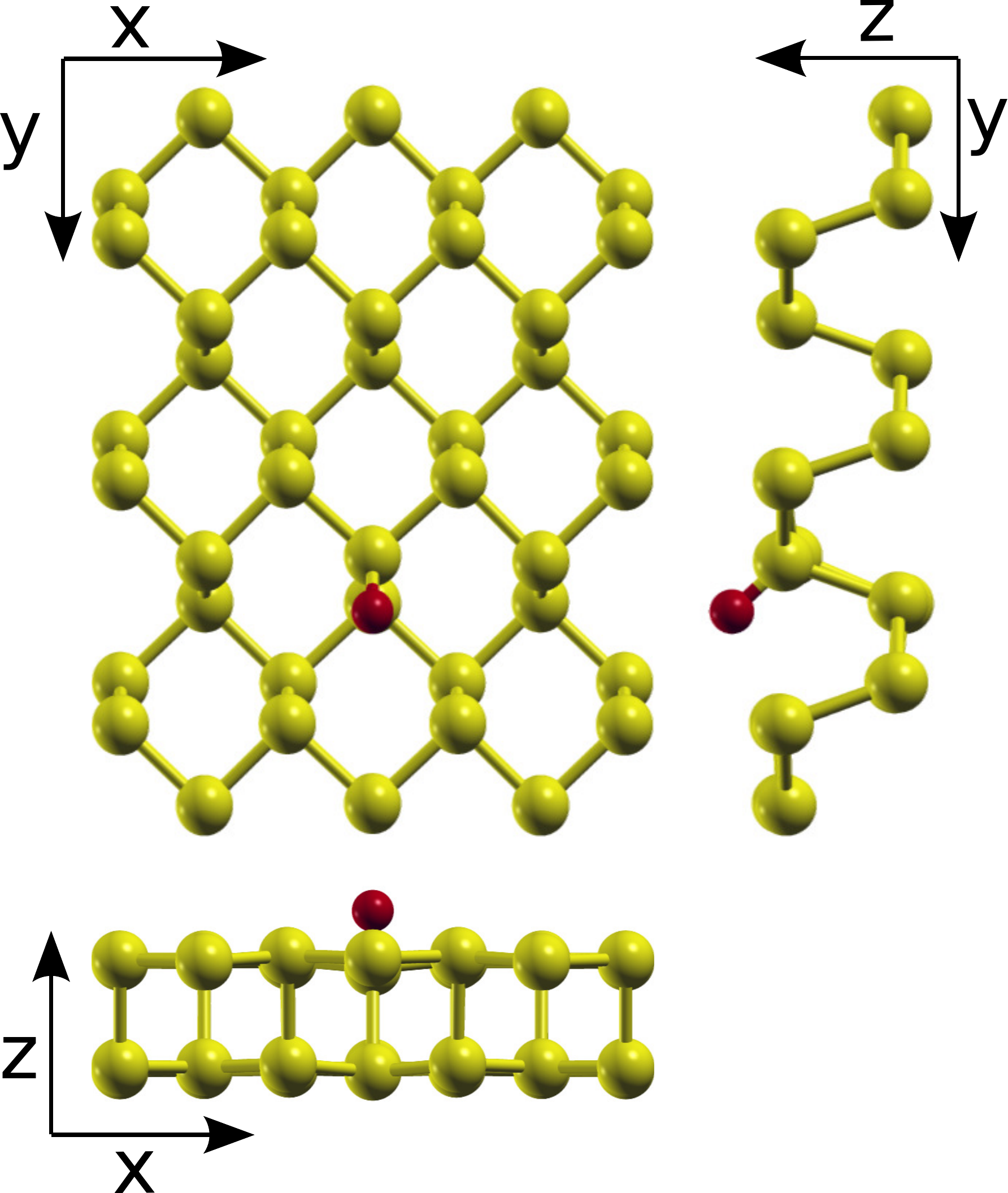}}
    \label{fig:dang1}
    \centering
      \subfloat[]{
    \includegraphics*[trim=30pt 25pt 0pt 0pt, width=5.5cm]{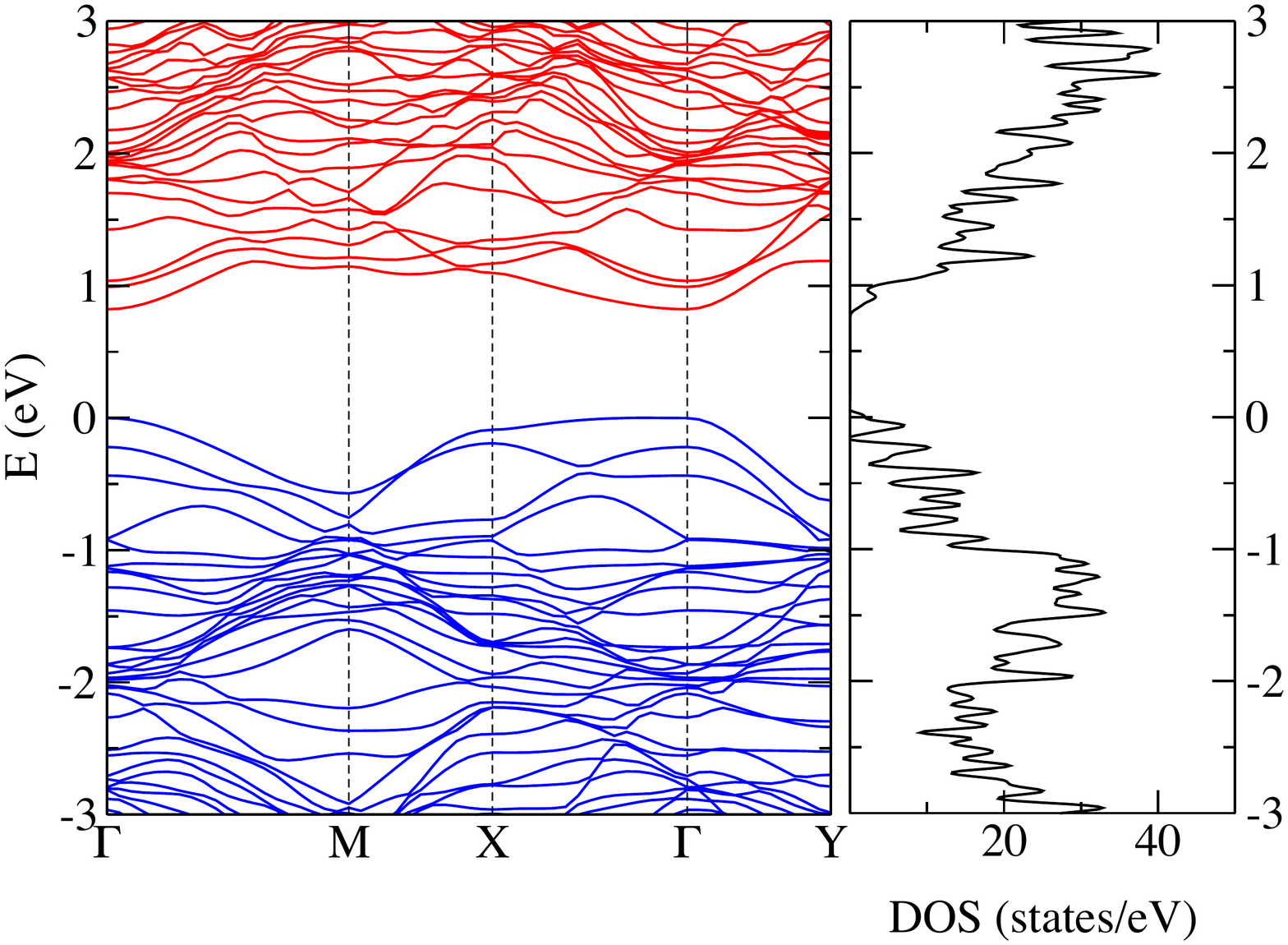}}
      \caption{ \small Dangling oxygen: (a) projections of the structure;
(b)  electronic band structure (left)  and DOS (right). The top of the valence band is set to zero.
The band structure of pristine phosphorene is provided in \emph{Supplemental Material}.
}\label{fig:dang}
\end{figure}

The lowest energy structure (highest binding energy) corresponds to an oxygen chemisorbed in a dangling configuration (Fig.\ref{fig:dang}a). 
The oxygen bonds with one phosphorus atom, with a P-O bond length of 1.50 \AA; 
the bond is short and polar, due to the large difference in electronegativity between P and O (2.19 ~eV vs 3.44~eV, respectively). 
Moreover, the P-O bond is tilted by $44.5^{\circ}$ away from the phosphorene surface.
The phosphorus involved in the P-O bond gets dragged into the lattice by 0.11~\AA\ in the $z$ direction; 
apart from that, the lattice deformation is minimal. 
The binding energy is very high, 2.08~eV and 1.82~eV at the PBE and HSE levels, respectively.  
Thus, the oxygen chemisorption is a strongly exothermic process, with a large energy gain as a result of the formation of the P-O bond,
even at the expense of an O=O bond. 
This defect is electrically neutral, as it introduces no states in the gap (Fig.~\ref{fig:dang}). The DOS (Fig.\ref{fig:dang}b) of the bands close to the Fermi energy is nearly identical to that of pristine phosphorene (see \emph{Supplemental Material}).

\begin{figure}[htb]
\centering
  \subfloat[]{%
    \includegraphics*[trim=0pt -40pt 0pt 0pt,width=2.9cm]{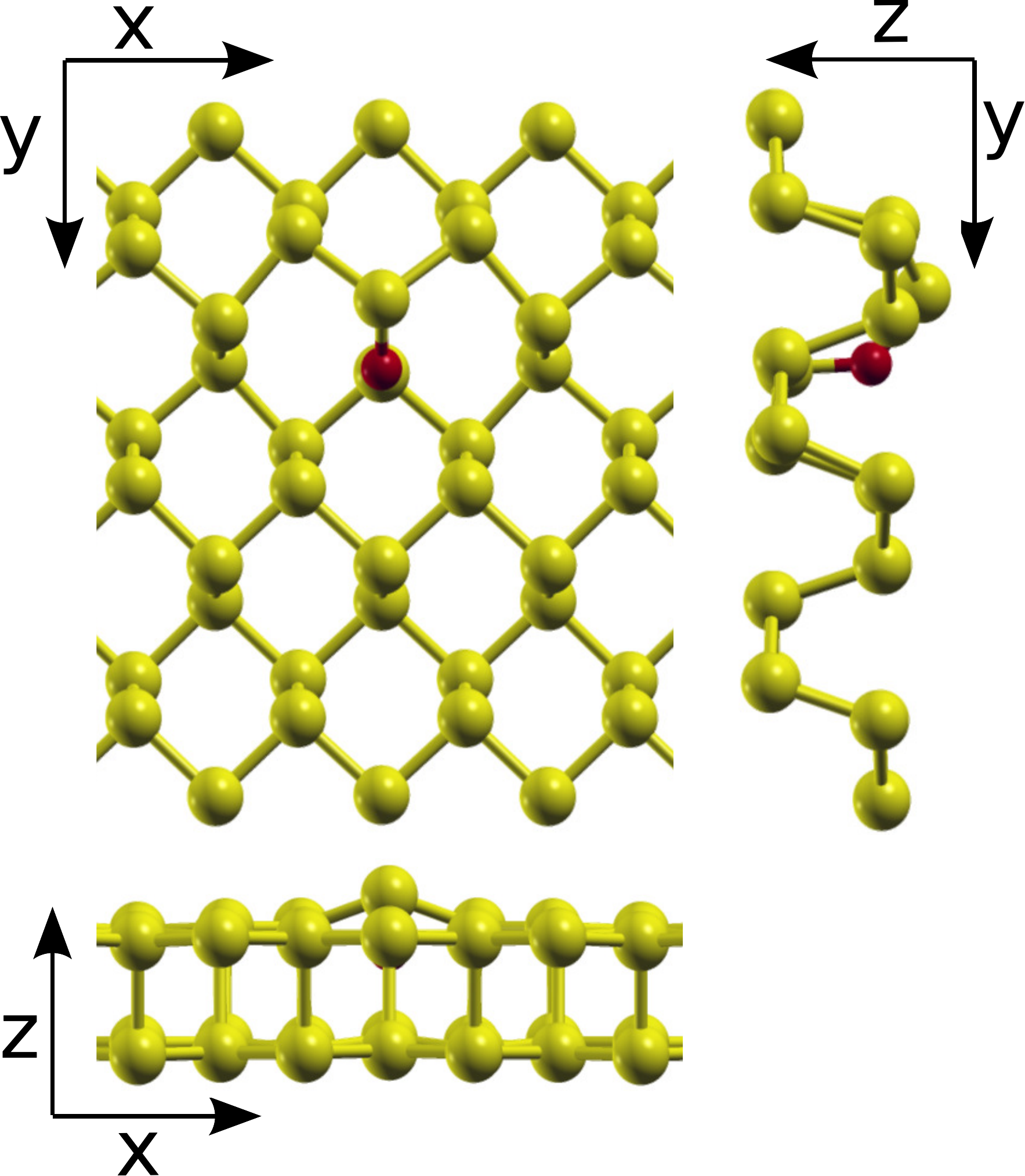}}
         \subfloat[]{%
    \includegraphics*[trim=0pt 20pt 0pt 80pt, width=5.5cm]{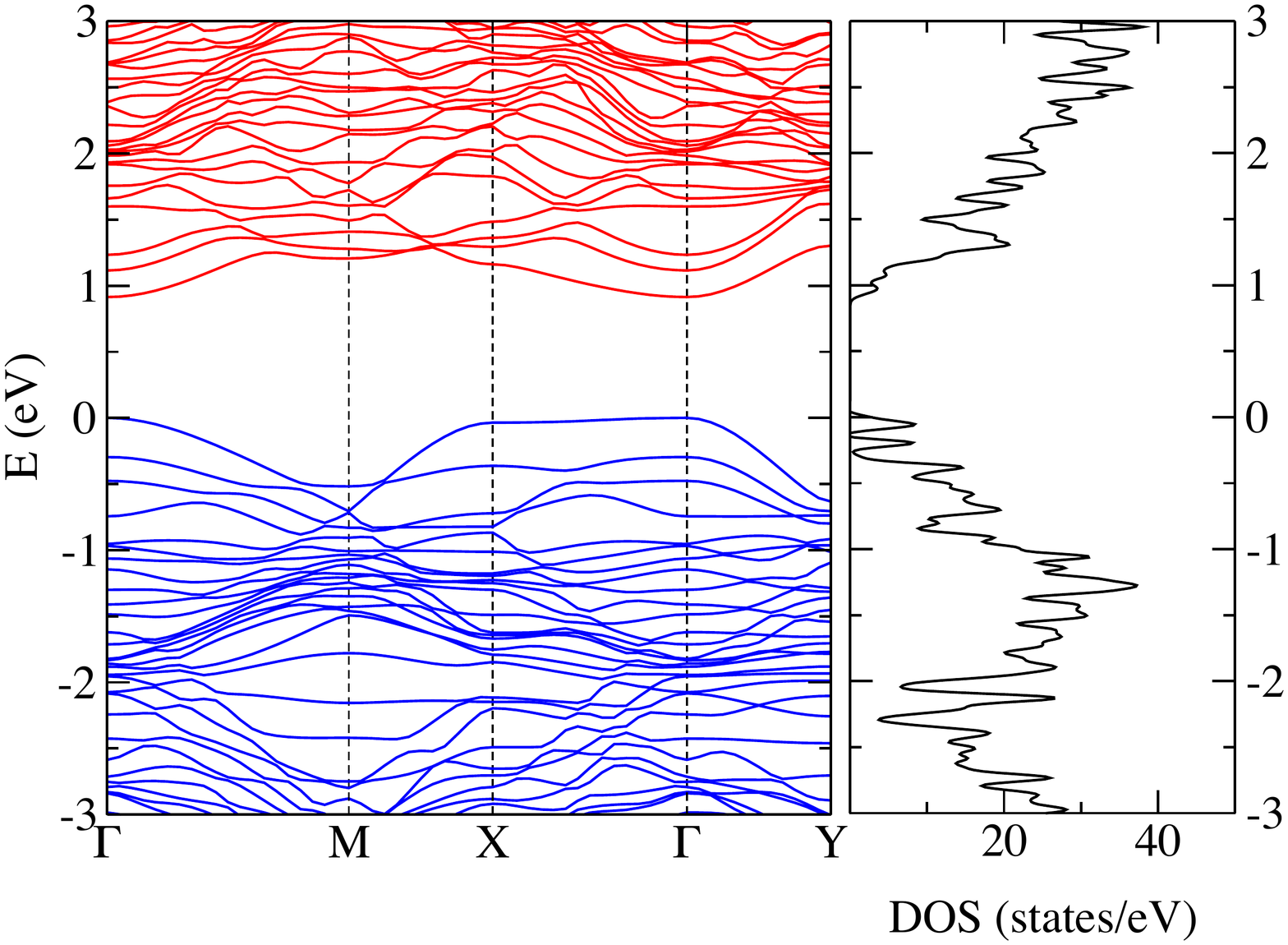}}
          \begin{minipage}{.1cm}
            \vfill
            \end{minipage}
  \caption{\small 
Interstitial oxygen: 
(a) projections of the structure; (b)  electronic band structure (left)  and DOS (right). The top of the valence band is set to zero.}  \label{fig:inter}
\end{figure}

The second lowest energy configuration, which is also electrically neutral,
is the interstitial oxygen bridge, shown in Fig.\ref{fig:inter}.  
In this case, the O penetrates into the lattice, occupying a position close to a P-P bond centre,
and forming a bridge between the two P atoms,
resembling the structure of interstitial oxygen in silicon\cite{coutinho2000}.
The resulting P-O bond lengths are 1.66~\AA\ and 1.68~\AA\,
 and the P-O-P bond angle is $129.6^{\circ}$. 
The oxygen atom pushes one of its P neighbors outwards, 
as clearly seen in Fig.\ref{fig:inter}a (bottom). 
The monolayer thickness is now 2.87~\AA, which corresponds to an increase of 36\% \emph{w.r.t.} pristine phosphorene (2.11~\AA). 
This huge deformation of the structure due to the oxygen bridge is likely
to contribute to the carrier scattering and even jeopardize the stability of the crystal, 
especially for larger impurity concentrations. 
Despite the large structural change of the lattice, the formation of interstitial oxygen bridges is highly exothermic, 
with a binding energy of 1.66~eV (1.60~eV) at the PBE (HSE) level. 
No states are formed in the middle of the gap (Fig.\ref{fig:inter}b), and the transformation from dangling to interstitial oxygen requires an activation energy of 0.69~eV. 

There are other metastable bridge-type surface defects with negative but small binding energies (Table \ref{table:e-bind}),
which may be formed if the oxygen source is more reactive than the O$_2$ ground state (for example, under light pumping).

\begin{figure}[htb]
\centering
  \subfloat[]{%
    \includegraphics*[trim=0pt -40pt 0pt 0pt,width=2.9cm]{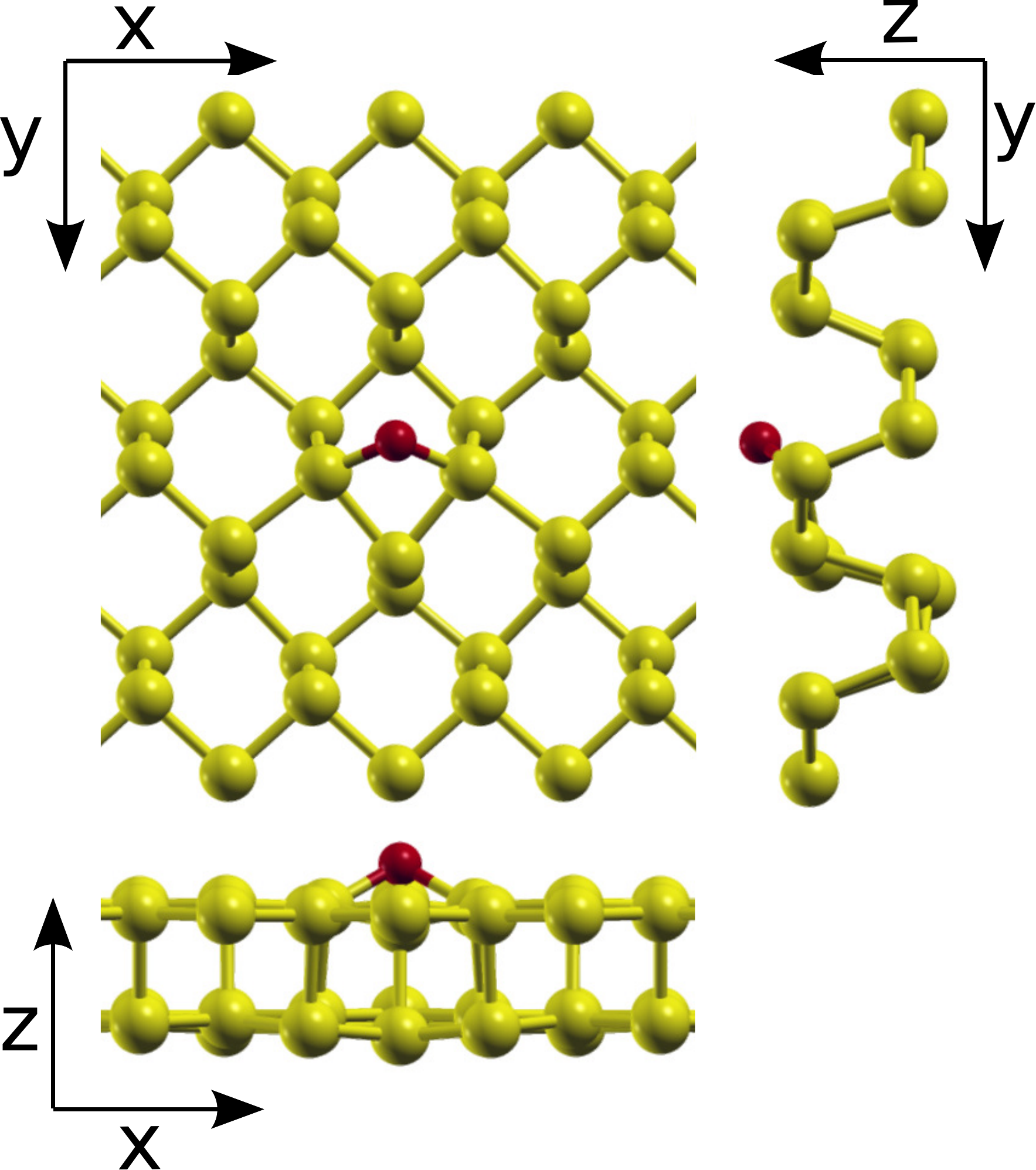}}
         \subfloat[]{%
    \includegraphics*[trim=0pt 20pt 0pt 80pt, width=5.5cm]{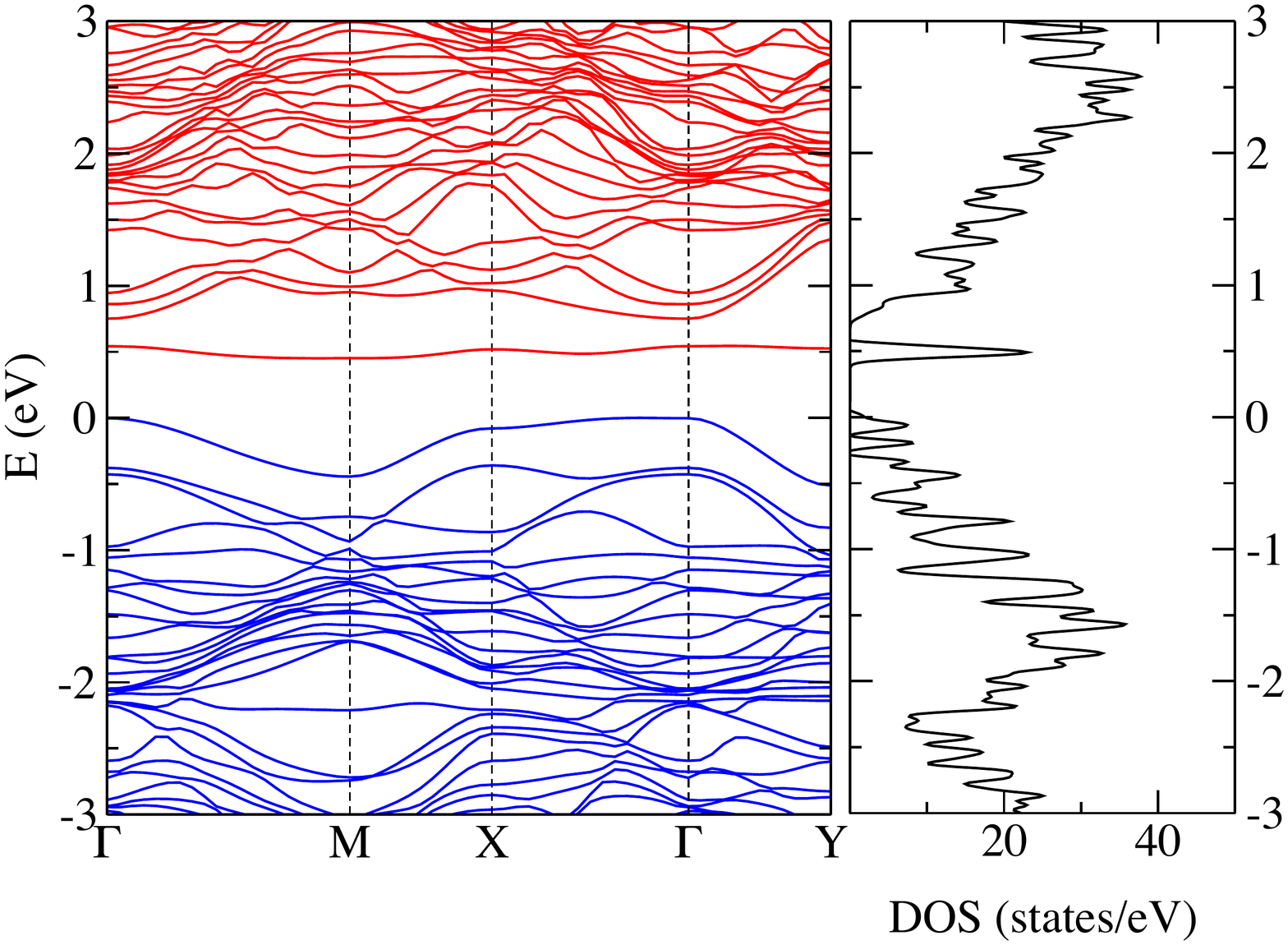}}
          \begin{minipage}{.1cm}
            \vfill
            \end{minipage}
  \caption{\small 
Horizontal oxygen bridge: 
(a) projections of the structure; (b)  electronic band structure (left)  and DOS (right). The top of the valence band is set to zero.}  \label{fig:bridge-h}
\end{figure}

Two possible configurations are found: the oxygen is on top of the zigzag ridges, and can either form a diagonal bridge (\emph{Supplemental Material}) -- connecting atoms on different edges of the zigzag -- or a horizontal bridge (Fig.\ref{fig:bridge-h}a) -- connecting atoms from the same edge. Both types of surface bridge defects create levels in the gap.

In the diagonal bridge configuration, the two P-O bonds are identical, with bond length of 1.73~\AA\ and bond angle of 77.0$^{\circ}$. 
The two P atoms involved in the bonding are dragged together by the oxygen bridge and their distance is reduced from 2.21~\AA\ in pristine phosphorene to 2.15~\AA.
This defect introduces an acceptor state and a perturbed valence band like state (see \emph{Supplemental Material}).
The perturbed valence band state is the result of the hybridization between $p_z$ orbitals of the oxygen and $p$ orbitals of neighboring P atoms, 
while the acceptor state consists of $p_x$ and $p_y$ orbitals from the O and $p$ orbitals from adjacent P atoms. 
The strong localization of the perturbed valence band state results in a large peak in the DOS.
By comparison, the acceptor state is slightly more dispersive, and thus produces 
a broader new set of states from 0 to 0.5~eV, just above the Fermi energy.

The horizontal bridge defect  has similar properties. The P-O bonds are now 1.75~\AA\ long and form an angle of 107.9$^{\circ}$.
The distance between the P atoms involved in the bonding is 2.83~\AA, decreased by 0.45~\AA\ \emph{w.r.t.} pristine phosphorene. 
This defect introduces a deep acceptor state at 0.5~eV near the conduction band.  
This mid-gap state is formed by the $p_x$ orbital of the O and the $p$ orbitals of the two P atoms involved in the P-O bonding. In addition, there is also a perturbed valence band state.
Such gap states are expected to give rise to recombination lines in luminescence experiments.

\begin{figure}[htb]
\centering
  \subfloat[]{%
    \includegraphics*[trim=0pt 0pt 0pt 0pt, width=3.5cm]{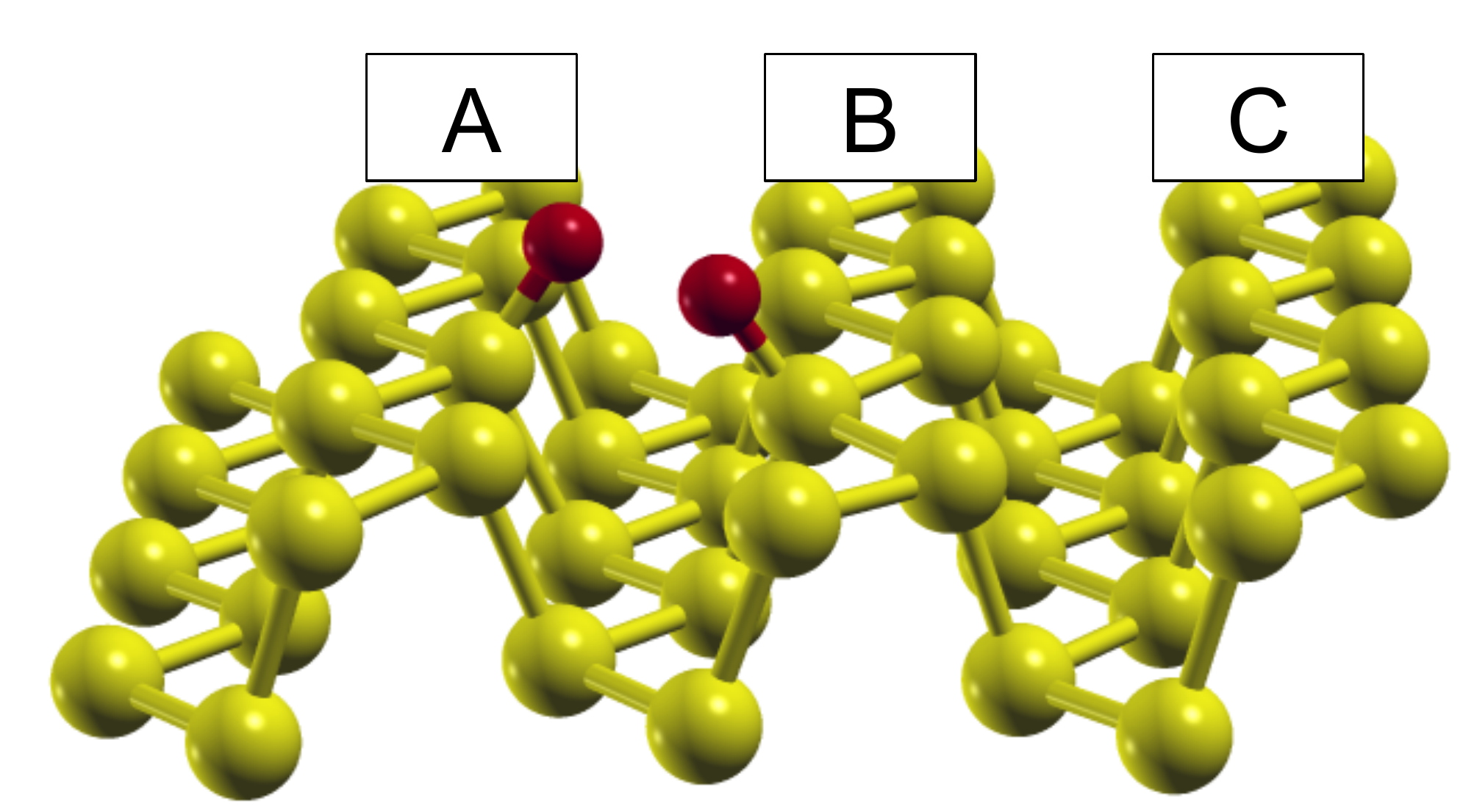}}
      \subfloat[]{%
    \hspace{15pt}
    \includegraphics*[trim=0pt 0pt 0pt 0pt, width=3.5cm]{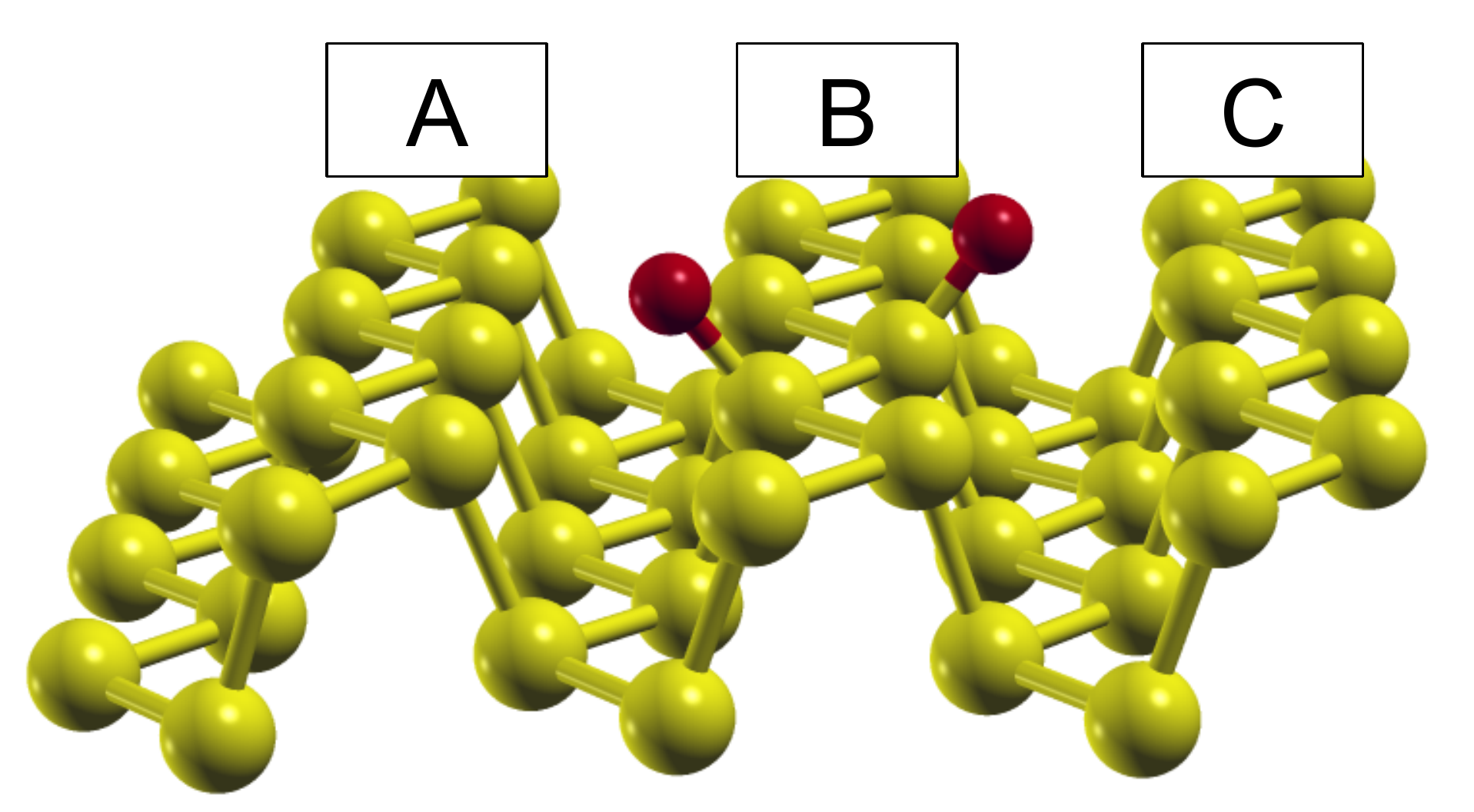}}
  \caption{\small  Possible configurations for two chemisorbed dangling oxygens. }  \label{fig:ox-dimer}
\end{figure}

Finally, for the most stable (dangling) configuration, we investigate the interaction of two chemisorbed oxygens. 
Chemisorbed oxygen atoms repel each other due to the Coloumb interaction between their $p$ orbitals. 
However, close configurations may be formed kinetically following the dissociation of oxygen molecules.
Even though the repulsion between neighboring oxygen atoms (e.g. Fig.\ref{fig:ox-dimer}a) reduces the binding energy per O atom of the complex by up to 0.4~eV, the dissociation of O$_2$ and subsequent formation of two P-O bonds is still very exoenergetic.
The oxygen atoms always point away from the zigzag ridge on which they are chemisorbed (labeled as A, B or C in Fig.\ref{fig:ox-dimer}), regardless the position of the other oxygen. 
The interaction between dangling oxygens depends on both distance and relative orientation of their P-O bonds. 
Configurations in which the oxygens point in different directions (e.g.  Fig.\ref{fig:ox-dimer}b) are favored \emph{w.r.t.} configurations where they point towards each other (e.g.  Fig.\ref{fig:ox-dimer}a).
For greater distances ($>$5 \AA), the 
electrostatic interaction between the P-O bond dipoles
and the strain interaction fall quickly,
resulting in repulsive energies below 0.01 eV. Thus, the chemisorption of oxygen molecules is likely to be a random process, 
or rather determined by the presence of local impurities or other defects. 

\begin{figure}[htb]
\centering
    \includegraphics*[width=8.2cm]{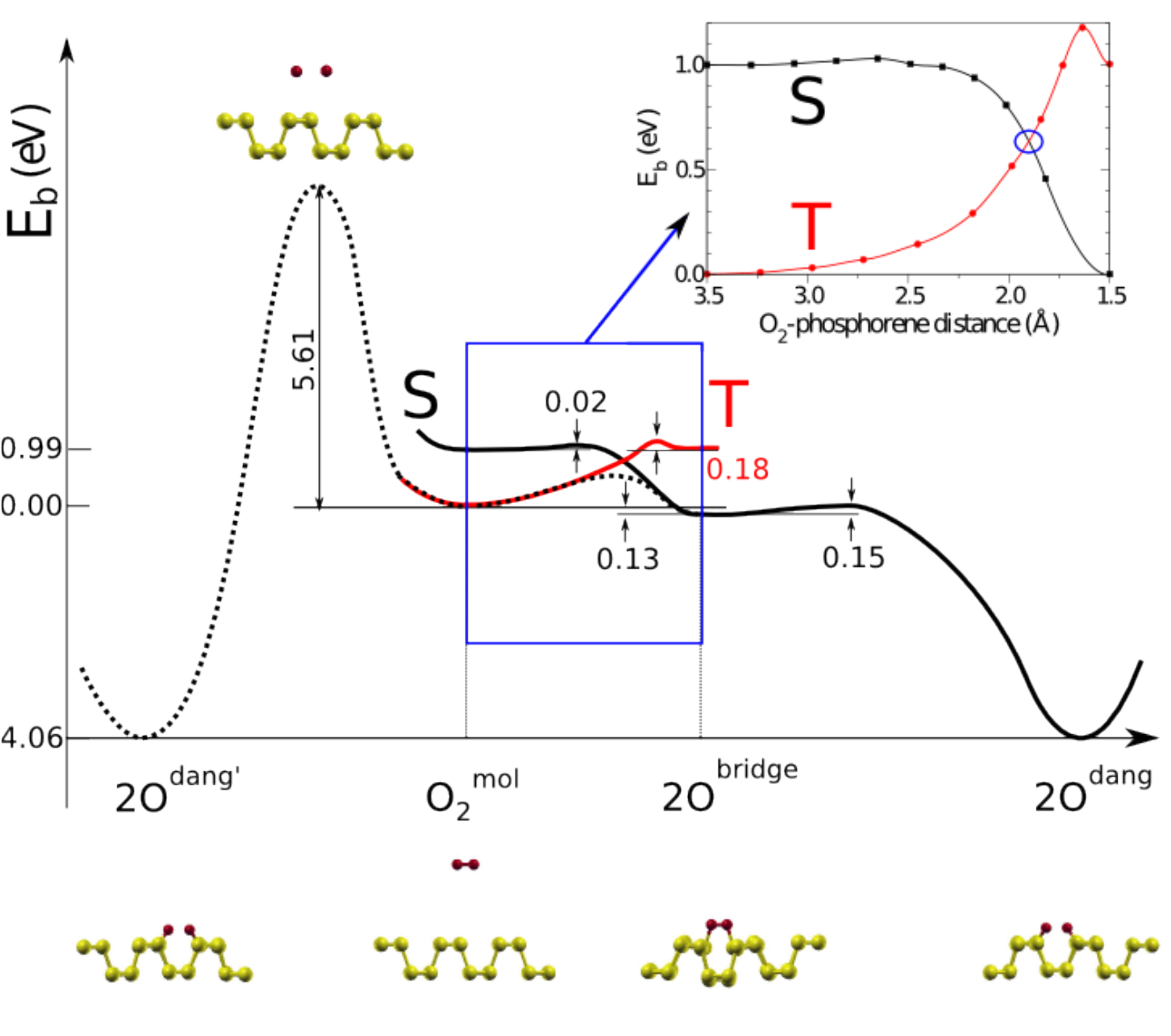}
 \caption{\small Schematic configuration-coordinate diagram for 
possible mechanisms for phosphorene oxidation. The solid (dotted) lines indicates the potential energy surfaces (PESs) calculated with fixed (variable) total magnetization.
Singlet (black) and triplet (red) PES as a function of the O$_2$-phosphorene distance are shown in the inset.}

\label{fig:mechanism}
\end{figure}

In a normal working environment, oxygen is present in its molecular form with its triplet ground state ($^3\Sigma_g^-$), in which the doubly degenerate $\pi_{2p}^*$ orbitals are each only half filled by two electrons with parallel spin. Dissociation of O$_2$, and subsequent formation of two dangling oxygens leads to an energy gain of more than 4~eV. 
Molecular oxygen can in principle dissociate on the phosphorene surface by either a direct or an indirect mechanism, 
through intermediate states where one or two oxygen atoms are bound to phosphorene,
as shown in Fig.\ref{fig:mechanism}. 
Direct dissociation of O$_2$, however, would require overcoming a barrier of 5.6~eV (Fig.\ref{fig:mechanism}, left side), close to its binding energy (6.2~eV at the PBE level\cite{O2-bind}) and therefore is highly improbable without photo excitation. 
Hence, an alternative indirect pathway is likely to be the main oxidation channel. 
Oxygen is initially physisorbed at 3.2~\AA\ parallel to the phosphorene surface with a small binding energy (0.01~eV and 0.08~eV at the PBE and PBE+vdW\cite{grimme2006} level, respectively). The system is in a triplet state, and no charge transfer between O$_2$ and phosphorene is observed. 
As the oxygen molecule gets closer to the surface,  the degeneracy of its $\pi_{2p}^*$ orbitals is lifted due to the hybridization with the phosphorene lone pairs, and the singlet state becomes more stable than the triplet. Therefore, there must be at least one point in configuration space where the triplet and singlet potential energy surfaces (PESs) cross, and triplet-to-singlet conversion likely takes place non-radiatively through intersystem crossing (ISC). This crossing occurs when the oxygen molecule is approximately 1.9~\AA\  above the phosphorene surface (see Fig.\ref{fig:mechanism}-inset).
Eventually O$_2$ is chemisorbed, forming a diagonal molecular bridge, with an energy gain of 0.13~eV \emph{w.r.t.} isolated (triplet) oxygen and pristine phosphorene. This is similar to one of the proposed pathways for oxidation of silicon\cite{fan2005,kato1998} and graphene \cite{zhou2013}.
As is clear from the PESs in Fig.\ref{fig:mechanism}, 
in the adiabatic approximation, reaching the oxygen molecular bridge configuration requires an activation energy that depends on the crossing point between triplet and singlet PESs. 
A spin-unrestricted variable-magnetization calculation gives a barrier of 0.54~eV. Once the oxygen bridge has been formed, however, only 0.15~eV is needed for the spin-allowed phonon-mediated dissociation of the O$_2$ bridge, and subsequent formation of two dangling O, with an energy gain of $\sim$3.9~eV. 
The bottleneck of phosphorene oxidation is then the initial chemisorption, where the system needs both to overcome an energy barrier and undergo an ISC, an inherently slow process. Landau-Zener\cite{zener1932,kato1998,orellana2003} theory provides an estimate of the probability $P_{ts}$ for the (single passage) triplet-to-singlet conversion: $P_{ts}=2\left[1-\exp(-V^2/hv\left|F_t-F_s\right|)\right]
\label{eq:p-ts}$,
where $V$ is the spin-orbit matrix element between the triplet and singlet states of free
O$_2$, $v$ is the velocity of an incident O$_2$ molecule, 
$F_s$ and $F_t$ are the slopes of the singlet and triplet PES at the crossing point, and $h$ is the Planck's constant.
By using $V=122$~cm$^{-1}$\cite{langhoff1974,kato1998,orellana2003}, and estimating $v$ from the O$_2$ center-of-mass thermal energy at 300~K\cite{orellana2003,zhou2013}, we obtain $P_{ts}=0.12$. Thus, the triplet-to-singlet conversion limits this oxidation channel. Nonetheless, due to the small energy barrier and the high exothermicity of the reaction, dangling oxygen defects will be present on the phosphorene surface after exposure to oxygen (or air). Using an attempt frequency $\nu=10^{13}$~s$^{-1}$, we estimate a rate of the order of $10^3$~s$^{-1}$ at room temperature.
In addition,  singlet oxygen is expected to readily oxidize phosphorene 
because only two low barriers (0.02 and 0.15~eV) separate the physisorbed oxygen from the lowest energy chemisorbed  configuration.

From these theoretical results, we can infer the effects of oxidation and how they can be detected.
The readiness with which phosphorene oxidizes in air is easily explained 
by the stability of the oxygen defects, which have binding energies of up to 2.1 eV per atom. 
However, the formation of surface dangling oxygen defects requires an activation energy of at least $\sim$0.54~eV 
leading to the structure shown in Fig.~\ref{fig:dang}(a). 
This energy has to be provided either thermally 
or possibly by light-induced excitation of the O$_{2p}$ electrons.
The penetration of O into the lattice from the dangling configuration requires an activation energy of 0.69~eV, and probably occurs in a subsequent stage.
The most stable oxygen defects are electrically neutral,
in analogy with the case of interstitial oxygen in Czocharlski silicon,
which in its isolated state leaves the electronic properties unaffected\cite{coutinho2000}.
Dangling oxygen defects cause little lattice deformation.
However, they change the surface affinity to interact with other species present in air.
Once some oxygen atoms are chemisorbed, the surface will be polarized, 
due to the difference in electronegativity between O and P, and will
become more hydrophilic.
Thus, even though dangling oxygen defects themselves do not change drastically the properties of the material,
they can be a starting point for the formation of more complex defects involving water 
or other polar molecules. The presence of oxygen defects might indeed play a role in the hydrophilicity of few-layer black phosphorus observed after air exposure\cite{gomez2014}.

Recently it has been claimed\cite{favron2014} that the simultaneous presence of oxygen, water and light is necessary for few-layer black phosphorus oxidation. Our results (Fig.\ref{fig:mechanism}) show that oxidation should also be possible in pure oxygen atmosphere (under light illumination). The issue seems to be how oxidation is detected. 
Conventional Raman spectroscopy is sometimes not sensitive to surface-adsorbed species on semiconductors, especially if present at low concentration\cite{quagliano2004}.
Moreover, since the experiment in Ref. [34] 
was performed on few-layer black phosphorus (not monolayer) the majority of the Raman signal probably comes from the pristine layers underneath the first oxidized layer; P-O vibrational modes might also be hidden beneath the SiO$_2$ substrate vibrational background.
Surface-sensitive spectroscopic techniques, such as surface-enhanced Raman spectroscopy\cite{sers1974} (SERS) or vibrational sum frequency generation\cite{vsfg2005} (VSFG), may be required for experimental detection and classification of oxygen defects.  We found that dangling oxygen gives rise to a local vibrational mode at 1099~cm$^{-1}$, while interstitial oxygen originates two local vibrational modes at 574~cm$^{-1}$ and 763~cm$^{-1}$.

For low oxygen concentrations, the activation energy for oxygen insertion (forming interstitial oxygen from an initial dangling configuration) is W= 0.69~eV. Estimating the transition rate $R=\nu\exp(-W/kT)$ using the typical vibrational attempt frequency $\nu=10^{13}$~s$^{-1}$,
the transition rate at room temperature is already of the order of 10~s$^{-1}$.
Thus, some interstitial oxygen may be formed at room temperature, albeit at a concentration orders of magnitude lower than dangling oxygen.

Nevertheless, oxygen insertion into the lattice results in considerable deformation, expanding the P-P distance, increasing the local layer thickness by 36\%.
This promotes further oxidation in the neighborhood of the pre-existing interstitial oxygen defects.
Further, for high enough interstitial oxygen concentration, the crystal might break apart due to this large stress, 
probably forming molecular compounds such as phosphorus trioxide ($\text{P}_4\text{O}_{6}$), 
phosphorus pentoxide ($\text{P}_4\text{O}_{10}$) or phosphates. 

The formation of other oxygen bridge-type defects 
is not energetically favored, but can be promoted under non-equilibrium
conditions, for example under light excitation.
Such defects are electrically active, introducing acceptor states and perturbed valence states.
It is possible that these acceptor states contribute to the $p$-type conductivity
observed in samples not intentionally doped\cite{shirotania1982,liu2014}.
However, in the case considered here of isolated defects in a monolayer, these defects
are localized and not shallow enough to confer high conductivity to the material.

Since the formation of the electrically active defects requires a larger activation energy,
a possible strategy to avoid their formation is to process the samples at low temperature.
Further, avoiding the contact of oxidized samples with water may prevent the formation of more complex defects and sample deterioration.
\\ \\ 
A.Z. and D.F.C. acknowledge NSF grant CHE-1301157 and also an allocation of computational resources from Boston University's Office of Information Technology and Scientific Computing and Visualization. A.Z. also acknowledges the support from NSF grant CMMI-1036460, Banco Santander. A.H.C.N. acknowledges NRF-CRP award ``Novel 2D materials with tailored properties: beyond graphene" (R-144-000-295-281). 

\bibliographystyle{unsrtnat}

\end{document}